\documentclass[aps,prl,twocolumn,amsmath,amssymb,nofootinbib,showpacs,superscriptaddress]{revtex4-1}
\usepackage[english]{babel}
\usepackage{latexsym}
\usepackage{graphics}
\usepackage{graphicx}
\usepackage{epsfig}
\usepackage{color}
\usepackage{bm}
\usepackage{amsmath}
\usepackage{amssymb}
\usepackage{amsthm}
\usepackage{dcolumn}
\usepackage{bm}
\usepackage{float}
\usepackage{color}
\usepackage{epstopdf}
\usepackage[svgnames]{xcolor}
\usepackage[colorlinks=true,citecolor=Cerulean,linkcolor=RubineRed,urlcolor=Cerulean,pdftex]{hyperref}
\hypersetup{hidelinks,colorlinks=true,allcolors=DarkBlue}
\usepackage{cleveref}

\usepackage{xcolor}
\definecolor{Cerulean}{rgb}{0.,0.59,0.835}
\definecolor{RubineRed}{rgb}{0.61,0.07,0.12}

\begin{document}

\preprint{APS/123-QED}

\title{Quantum phase transition of a two-dimensional quadrupolar system}

\author{G.E. Astrakharchik}
\affiliation{Departament de F{\'i}sica, Campus Nord B4-B5, Universitat Polit{\`e}cnica de Catalunya, E-08034 Barcelona, Spain}

\author{I.L. Kurbakov}
\affiliation{Institute of Spectroscopy, Russian Academy of Sciences, Troitsk, Moscow 142190, Russia}

\author{D.V. Sychev}
\affiliation{School of Electrical and Computer Engineering, Purdue University, West Lafayette, Indiana 47907, USA}

\author{A.K. Fedorov}
\affiliation{Russian Quantum Center, Skolkovo, Moscow 143025, Russia}
\affiliation{Moscow Institute of Physics and Technology, Dolgoprudny, Moscow Region 141700, Russia}

\author{Yu.E. Lozovik}
\affiliation{Institute of Spectroscopy, Russian Academy of Sciences, Troitsk, Moscow 142190, Russia}
\affiliation{Center for Basic Research, All-Russia Research Institute of Automatics, Moscow 127055, Russia}
\affiliation{MIEM at National Research University HSE, Moscow 109028, Russia}

\date{\today}
\begin{abstract}
Ensembles with long-range interactions between particles are promising for revealing strong quantum collective effects and many-body phenomena.  
Here we study the ground-state phase diagram of a two-dimensional Bose system with quadrupolar interactions using a diffusion Monte Carlo technique.
We predict a quantum phase transition from a gas to a solid phase.
The Lindemann ratio and the condensate fraction at the transition point are $\gamma=0.269(4)$ and $n_0/n=0.031(4)$, correspondingly. 
We observe the strong rotonization of the collective excitation branch in the vicinity of the phase transition point.
Our results can be probed using state-of-the-art experimental systems of various nature, such as quasi-two-dimensional systems of quadrupolar excitons in transition metal dichalcogenide (TMD) trilayers, quadrupolar molecules, and excitons or Rydberg atoms with quadrupole moments induced by strong magnetic fields.
\end{abstract}
\maketitle

Manipulation of the short-range interactions in ultracold quantum gases has proven to be an efficient and productive way to generate novel many-body phases~\cite{Bloch2008,Bloch2012,Bloch2017}. 
Even more evolved scenarios are realized in gases with long-range interactions such as dipolar ones~\cite{Baranov2008,Pfau2009,Baranov2012}.
Dipolar particles interact with each other via anisotropic and long-range forces, which drastically changes the structure of many-body phases in these systems both in the free space and lattices~\cite{Baranov2008,Pfau2009,Lewenstein2011,Baranov2012}.
Remarkable progress in experiments with ultracold gases of large-spin atoms~\cite{Pfau2005,Lev2011,Lev2012,Ferlaino2012} and polar molecules~\cite{Ye2009,Gabbanini2009,JinYe2016} has opened up fascinating prospects for the experimental observation of novel quantum phases,
which are induced by the character of the dipolar interaction.
Examples include, in particular, 
rotonization~\cite{Shlyapnikov2003,Wilson2008,WilsonTicknor2012,Shlyapnikov2013,Fedorov2014},
crystallization~\cite{Lozovik2007,PupilloZoller2007,Boning2011}, and 
supersolidity for both dilute~\cite{Shlyapnikov2015,Ferlaino2018,Ferlaino2019,Pfau2017,Ferlaino20192,Modugno2019,Pfau2019,Boninsegni2017,Blakie2018,Roccuzzo2019} 
and dense~\cite{Pupillo2010,Pohl2010,Lozovik2011,PupilloZoller2007,Lozovik2007,Kurbakov2010,Boninsegi2011} dipolar systems. 
However, the interactions between atomic dipoles are typically weak.
This fact has stimulated the exploration of novel platforms with both strong interparticle interactions and sufficient tunability.
Examples include long-lived excitons in solid-state systems~\cite{Lozovik1975,Shevchenko1994,MacDonald2004,Timofeev2006,Snoke2010,Butov2012,Rapaport2014,Voronova2018}.
Remarkable advances in experiments with monolayers of semiconducting transition metal dichalcogenides (TMDs)~\cite{Shan2016,Geim2013} make them interesting for revealing non-conventional quantum phenomena~\cite{Novoselov2014,Rivera2015,Geim2016,Wurstbauer2017,Geim2018} in the regimes that are beyond what can be achieved with ultracold gases. 
TMD systems host long-lived excitons since the overlap between wavefunctions of electrons and holes locating in separate layers is suppressed, and the separation results in the appearance of the exciton dipole moment as it was predicted~\cite{Lozovik1975,Shevchenko1994}.
Dipolar excitons in solid-state systems might manifest rotonization~\cite{Lozovik2007,Kurbakov2010,Fedorov20142,Lukin2020} and supersolidity~\cite{Kurbakov2010}. 

Quadrupolar interactions present a peculiar example of non-local interactions between particles~\cite{Lemeshko2013,MatheyLemeshko2014,MatheyLemeshko2015}, which can be fine-tuned using external fields. 
This makes quadrupolar systems a promising platform for performing the quantum simulation and revealing novel many-body phases and unconventional quantum states~\cite{Lemeshko2013,MatheyLemeshko2014,MatheyLemeshko2015}.
Experimental realizations of quadrupolar ensembles include quadrupolar molecules, whose interaction is induced and tuned by external fields, and quadrupolar excitons in solid-state systems.
Quadrupolar species of particles acquiring electric quadrupole moments, such as Cs$_2$~\cite{Grimm2003} or Sr$_2$~\cite{Schreck2012,Zelevinsky2012}, are available in experiments.
Moreover they are stable against collapse and ultracold chemical reactions at high densities, which are shortcomings for experiments with dipolar molecules~\cite{Ye2009,Gabbanini2009,JinYe2016}.
Recent studies of TMD systems~\cite{Rapaport2020} have shown the rich many-body physics, which is induced by the nature of quadrupolar interactions.
We also note that classical quadrupolar interactions arise in soft matter in the description of nematic colloids. 
Their properties, including phase transitions, have been extensively studied in Refs.~\cite{Stark2001,Polson1997,Mondain-Monval1999,Musevic2006,Skarabot2008,Pergamenshchik2010,Ognysta2011}.
Besides, quadrupoles play an essential role in astrophysical objects in ultrastrong magnetic fields, e.g., on the surface of neutron stars~\cite{prl027001306}.
However, a detailed microscopic study and {\it ab initio} simulations of the quadrupole many-body system are still lacking.

Here we predict a quantum phase transition from a gas to a crystal in a single-component two-dimensional (2D) Bose system with centrally symmetric quadrupolar interactions at zero temperature.
We employ a diffusion Monte Carlo (DMC) technique for calculating the parameters of the phase transition and to study the effects of strong correlations in the gas phase.
Our results are in the quantitative agreement with predictions based on the quantum hydrodynamic (HD) model. 
We observe a roton-maxon character of the collective excitation branch.
The predicted results can be probed in state-of-the-art experiments with ultracold atoms (e.g., Rydberg atoms), molecular ensembles and TMD systems. 

\begin{figure}
\includegraphics[width=\columnwidth]{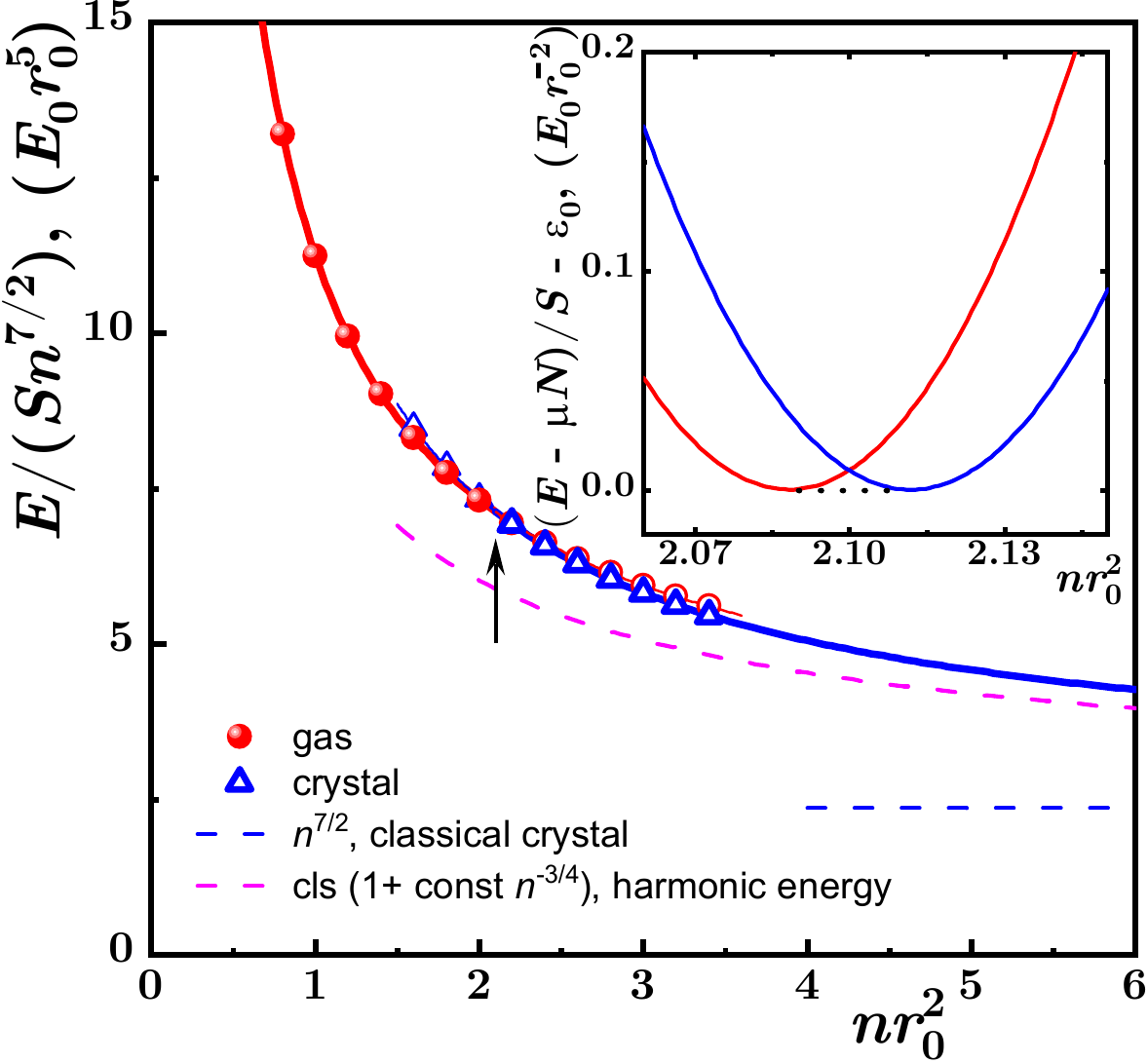}
\vskip -4mm
\caption{
The macroscopic limit of the energy $E/S$ (scaled with classical $n^{7/2}$ dependence) for the gas (circles) and the crystal (triangles) as a function of the dimensionless density $nr_0^2$ (the energy $E$ is measured in the dimensionless units $\hbar^2/mr_0^2$).
The position of the transition point, $nr_0^2=2.10(7)$, is indicated by the arrow.
Inset: the quantity $(E-\mu N)/S-\varepsilon_0$ as a function of the dimensionless density in the vicinity of the phase transition where $\varepsilon_0$ is an offset.
The tangent dotted line indicates the coexistence regime of two phases, its width is $\Delta nr_0^2=0.026(4)$.
The fitting function is $E/(NE_0)=E_{\rm cls}/(NE_0)+A_1(nr_0^2)^{7/4}+A_2(nr_0^2)^{5/4}+A_3(nr_0^2)^{3/4}$. 
Fitting coefficients are $A_1=7.944$, $A_2=-0.388$, $A_3=1.332$ for gas at $0.8<nr_0^2<3$ and $A_1=6.1478$, $A_2=2.4524$, $A_3=0.9878$ for crystal at $1.6<nr_0^2<3.4$, where $E_{cls}/(NE_0)=A_0(nr_0^2)^{5/2}$ with $A_0=2.359746$ is the ground-state energy of a classical crystal.}
\label{fig:energy}
\end{figure}

The Hamiltonian of a homogeneous system of $N$ bosons with the quadrupolar interaction is as follows:
\begin{eqnarray}\label{eq:Hamiltonian}
\mathcal{H}=-\frac{\hbar^2}{2m}\sum_{i=1}^{N}{\Delta_i}+\frac{Q^2}{\epsilon}\sum_{j<k}^{N}{\frac{1}{|{\bf r}_j-{\bf r}_k|^5}},
\end{eqnarray}
where $m$ is the particle mass, ${\bf r}_i$ is the 2D position of $i$-th particle, $Q$ is the quadrupolar moment and $\epsilon$ is the dielectric constant. 

It is convenient to rewrite Hamiltonian~(\ref{eq:Hamiltonian}) in a dimensionless form by expressing all the distances in units of $r_0=\sqrt[3]{mQ^2/\hbar^2\epsilon}$ and energies in units of $E_0=\hbar^2/mr_0^2$.
The characteristic quadrupolar length $r_0$ is directly proportional to the quadrupole-quadrupole $s$-wave scattering length, $a_s/r_0{=}(e^{\gamma_E}/3)^{2/3}=0.706383$ with $\gamma_E=0.577\dots$ the Euler constant.
We calculate the zero-temperature phase diagram of the system in terms of the dimensionless density $nr_0^2$, where $n$ is the 2D density of the system. 

In order to find the system properties we resort to the DMC technique~\cite{Boronat1994} based on solving the Schr{\"o}dinger equation in imaginary time and allowing one to obtain the exact ground-state energy.
The convergence is significantly improved by using an importance sampling for which we chose the trial wave function in the Nosanow-Jastrow product form~\cite{Lozovik2007}. 
Using the standard prescription, each particle in the solid phase is localized close to its lattice site by a one-body Gaussian term of variable width. 
An infinite width is used in the gas phase which results in a wave function having translational invariance.  
We chose the two-body Jastrow term as
\begin{equation}\label{Jastrow}
f_2(x)=\left\{\begin{array}{ll}
C_1K_0(2x^{-3/2}/3),\;&x\le x_c,\\
C_2\exp(-C_3/x-C_3/(\bar L-x)),\;&x_c\le x\le\bar L/2,\\
1,\;&\bar L/2\le x,
\end{array}\right.
\end{equation}
%where $x{=}r/r_0$, $\bar L=L/r_0$ and $x_c$ is the variational parameter 
where $x{=}r/r_0$, $\bar L=L/r_0$, $L$ is the length of the smallest side of the simulation box and $x_c$ is the variational parameter (matching point between the two-body scattering solution at short distances and the phononic long-range decay~\cite{pr0155000088}).
Coefficients $C_1,C_2,C_3$ are fixed by the condition of the continuity of the function and its first derivative.

The thermodynamic limit is then reached by increasing the number of particles while keeping the density $n=N/(L_x{\times}L_y)$ fixed and performing extrapolation to $N\to\infty$~\cite{comment,Comment1}.
We simulate systems containing $N=100, 144, 256, 484$ and $1156$ particles in a simulation box of size ${L_x}\times{L_y}$ with periodic boundary conditions. 
We use a square box with equal sides ${L_x}={L_y}$ for simulation of the gas phase and a rectangular box commensurate with an elementary cell of a triangular lattice for the solid phase.

\begin{figure}
\includegraphics[width=\columnwidth]{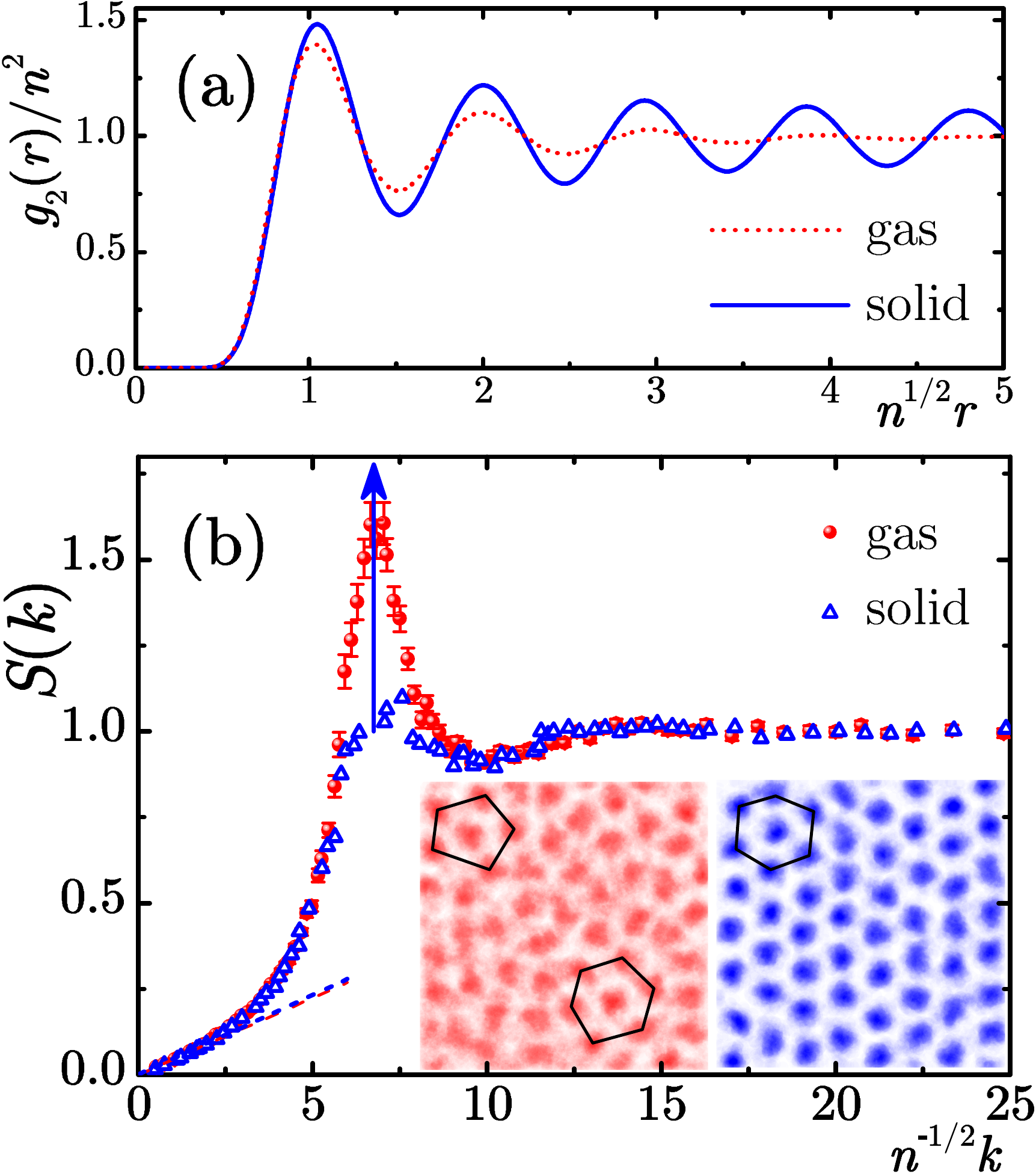}
\vskip -4mm
\caption{(a) Typical examples of the pair correlation function in gas (dashed line) and solid (solid line) phases at the density $nr_0^2=2.2$ obtained for $N=144$ particles. (b) Static structure factor in the vicinity of the phase transition in gas (circles) and triangular solid (triangles) phases.
Symbols, DMC data; 
lines, linear phonons asymptotic $S(k) = \hbar k /(2mc)$ where the speed of sound $c = \sqrt{n/m~d^2(E/S)/dn^2}$ is obtained from the equation of state, see Fig.~\ref{fig:energy}.
Vertical arrow shows the position of the macroscopic peak in the crystal.
Insets show snapshots of the particles' coordinates in gas (left) and solid phases (right). 
Polygons indicate a frustrated (left) and perfect (right) hexagonal short-range ordering present in gas and solid phases, correspondingly.
}
\label{fig:pair}
\end{figure}

We find a quantum phase transition from a gas to a solid phase at zero temperature.
In order to demonstrate its presence, we calculate the lowest energy in a state with translational symmetry (i.e. gas) and a state in which the translational symmetry is broken (i.e. solid). 
The phase transition density is obtained as the crossing between the energies of the two states. 
The two equations of state are shown in Fig.~\ref{fig:energy}, where for convenience the energy $E$ is scaled with the system area $S$ and density as $(E/S)/n^{7/2}$.
For small values of the dimensionless density $nr_0^2$ the energetically favorable state is a gas whereas the solid phase remains metastable.
As the density is increased, the system experiences a first-order quantum phase transition to a triangular lattice phase. 
We estimate the transition density to be $nr_0^2=2.10(7)$ (see Fig.~\ref{fig:energy}) with the width of the coexistence of the phases $\Delta nr_0^2=0.026(4)$.
There are remarkable differences as compared to a dipolar system which has a significantly larger critical density, $nr_{dd}^2\approx 290$ in dipolar units $r_{dd}=3.17a_s$\cite{Lozovik2007}. 
The difference becomes even more evident in terms of the $s$-wave scattering length, as the critical density is $na_s^2 \approx 1.05$ for quadrupoles, $na_s^2=2900$ for dipoles and $na_s \approx 0.33$ for hard disks\cite{Xing1990}. 

The Lindemann ratio quantifies the fluctuations of particles in a crystal and is defined as follows
\begin{equation}\label{gammaL}
\gamma=\sqrt{\sum\nolimits_{i=1}^N
\langle({\bf r}_i-{\bf r}_i^{\rm latt})^2/b^2\rangle},
\end{equation}
where $b=(4/3)^{1/4}/\sqrt{n}$ is the triangular lattice period. 
We find the Lindemann ratio to be $\gamma=0.269(4)$ at the transition point. 
In the limit of high density, the potential energy dominates and the energy gradually approaches that of a perfect classical crystal corresponding to the horizontal line in Fig.~\ref{fig:energy}.
For comparison, we also show in Fig.~\ref{fig:energy} the first correction to the classical crystal energy arising from the zero-point motion in harmonic approximation, $E/S=E_{\rm cls}/S+An^{11/4}$. 

In order to quantify the two-body correlations we calculate the pair distribution function,
\begin{equation}\label{g2}
g_2(r)\!=\!\int\limits_0^{2\pi}\!\frac{d\varphi}{2\pi}
\int\limits_0^{L_x}\!\int\limits_0^{L_y}\!\frac{d{\bf s}}{L_xL_y}\langle\hat\Psi^+({\bf s})\hat\Psi^+({\bf r\!+\!s})
\hat\Psi({\bf r\!+\!s})\hat\Psi({\bf s})\rangle,
\end{equation}
where $\varphi$ is the polar angle of the vector ${\bf r}$.
We show characteristic examples in Fig.~\ref{fig:pair}. 
Close to the transition point, the short-range correlations are very similar in both phases (see Fig.~\ref{fig:pair}a for separations smaller than the mean interparticle distance). 
Instead there are qualitative differences for larger separations $r$. 
In the gas phase, $g_2(r)$ approaches a constant value already after a few oscillations.
Instead, the oscillations continue further in the solid phase, signaling the presence of the diagonal long-range order.

The order parameter differentiating two phases is the height of the peak in the static structure factor 
\begin{equation}
S({\bf k})=\int\langle\hat\rho({\bf r})\hat\rho({\bf s})\rangle
e^{i{\bf k(r-s)}}d{\bf r}d{\bf s}/N,  
\end{equation}
at the reciprocal lattice period $k_L=2\pi\sqrt n(4/3)^{1/4}$ of the triangular crystal, where $\hat\rho({\bf r})$ is the density operator and $\langle \cdots\rangle$ denotes ground state averaging.
The characteristic feature of a crystalline phase is that the value of $S(k_L)$ is linearly proportional to the number of particles and the peak becomes macroscopic in the thermodynamic limit. 
This should be contrasted to the behavior in the gas phase in which the static structure factor always remains finite, see Fig.~\ref{fig:pair} for characteristic examples.
In that case, $S(k)$ is a monotonous function of momentum at low densities and it becomes non-monotonous (i.e. a peak is formed) in the regime of strong quantum correlations.
The height of the peak increases as the density is incremented and the phase transition from the gas to the crystal happens when the critical value, $ S(k)_{\rm max}=1.6(1)$, is reached.
There is a discontinuity in the order parameter, $S(k_L)$, across the phase transition point which is typical behavior for the first-order phase transition.
At the same time, the low-momentum behavior, $S(k)=\hbar k /(2mc)$, is more similar in the two phases which reflects a relatively minor change of the speed of sound $c$ across the transition [
compare two dashed straight lines at small momenta in Fig.~\ref{fig:pair}b].

The appearance of the short-range ordering in the gas phase in the vicinity of the critical density can be seen from the snapshots shown in the inset of Fig.~\ref{fig:pair}b.
The snapshot of the gas phase indicates the formation of a local triangular lattice with vacancies and dislocations, whereas a defect-free triangular lattice is observed in the ground state of the solid phase.

\begin{figure}
\includegraphics[width=0.9\columnwidth]{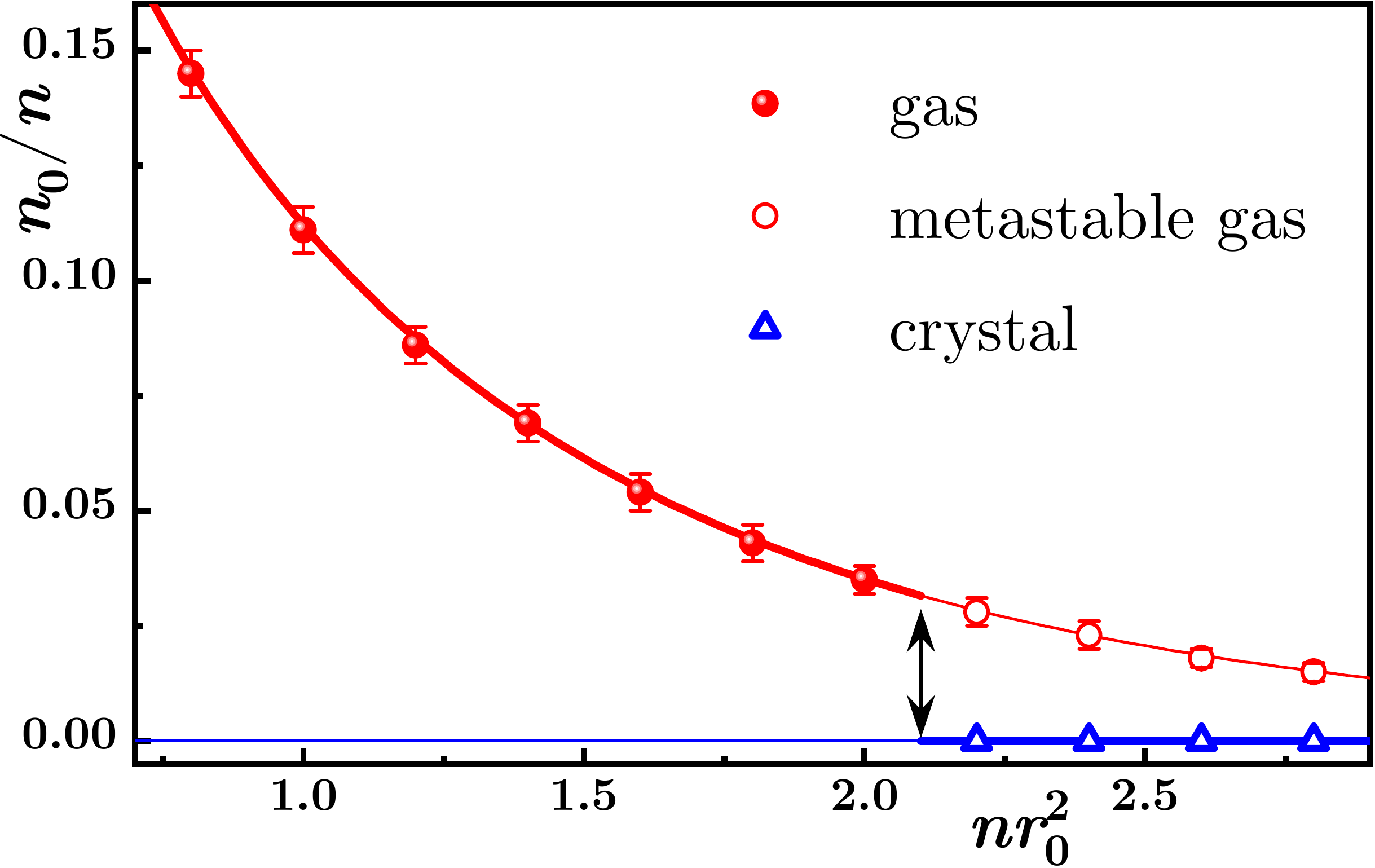}
\vskip -3mm
\caption{The condensate fraction $n_0/n$ in the macroscopic system as the function of the density in gas and solid phases.
Circles, extrapolation of Quantum Monte Carlo data to thermodynamic limit performed by using hydrodynamic theory [QMC+HD: input $S(k)\& g_1(L/2)$] of Ref.~\cite{BoronatPRL};
red line, fit $n_0/n=\exp\left[-(B_0+B_1(nr_0^2)^{B_2})^{-2}/4\right]$
in the region ${0.8}\leq{n}\leq{2.8}$, where $B_0=-0.301$, $B_1=0.639$, and $B_2=-0.154$.
The discontinuity at the phase transition is shown with arrows.}
\label{fig:OBDM}
\end{figure}

The coherence properties are quantified by the condensate fraction which is reported in Fig.~\ref{fig:OBDM}. 
We have verified that in a finite-size system, the long-range behavior of the one-body density matrix (OBDM) $g_1(r)$ is well reproduced by the hydrodynamic theory of Ref.~\cite{BoronatPRL}. 
We use the HD theory for the extrapolation of the OBDM in order to obtain the condensate fraction according to $n_0/n = \lim_{r\to\infty}g_1(r)$.  
We observe a strong condensate depletion as the density $nr_0^2$ is increased, so the value $n_0/n$ becomes fairly small close to the gas-solid transition. 
%Even if the value of a few percent might seem very small, conceptually it is important that $n_0/n$ experiences a sudden jump to zero value in the solid phase. 
Even if the condensate fraction is small, $n_0/n$ = 0.02 -- 0.04 conceptually it is important that it experiences a sudden discontinuous jump from a finite value in the gas phase to zero value in the solid phase.
In other words, the condensate fraction is another order parameter and together with the height of the structure factor both order parameters are discontinuous across the first-order phase transition. 
Both order parameters have exactly the same critical point and as a consequence a supersolid (simultaneous presence of both order parameters) is absent in the present system.

\begin{figure}
\includegraphics[width=0.35\textwidth]{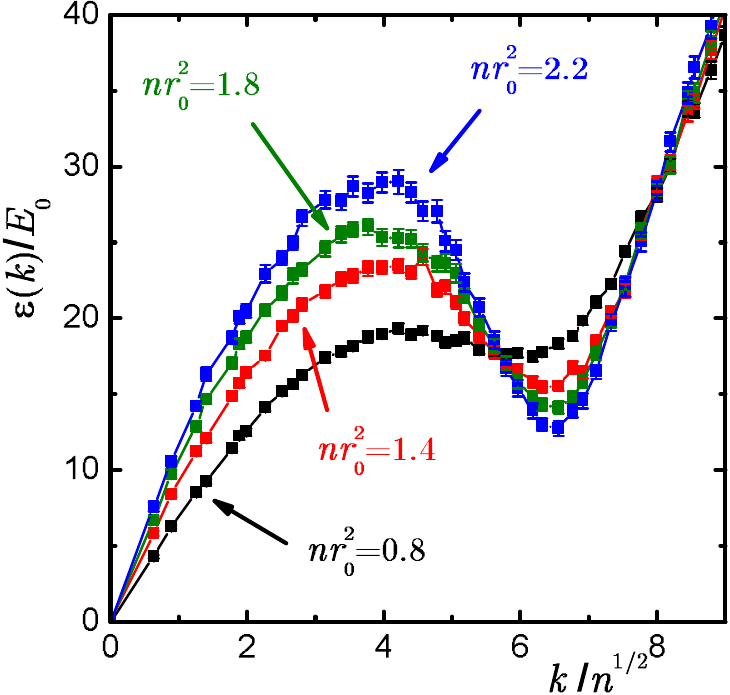}
\caption{Characteristic examples of the excitation spectrum in the gas phase as obtained from Feynman relation. 
The formation of a pronounced roton minimum is observed as density is increased and the transition to the solid phase is approached.}
\label{fig:spectrum}
\end{figure}

The rotonization of the excitation spectrum is yet another non-trivial effect that might be present in strongly-correlated systems and it deserves special attention~\cite{Shlyapnikov2003,Wilson2008,WilsonTicknor2012,Shlyapnikov2013,Fedorov2014}. 
Indeed, rotonization of the collective excitation branch may potentially lead to the spontaneous establishing of crystalline order and formation of supersolid. 
One of the possible mechanisms for supersolidity of dipolar systems is its formation near the gas-solid phase transition~\cite{Lozovik2007,PupilloZoller2007,Kurbakov2010}. 
Here we make evident the rotonization of the spectrum in the quadrupolar system by analyzing the Feynman relation which provides the upper bound for the lowest border of the excitation spectrum,
\begin{eqnarray}\label{eq:spectrum}
\varepsilon_k =\frac{\hbar^2k^2}{2mS(k)}, 
\end{eqnarray}
in terms of the static structure factor $S(k)$. 
Our results for the excitation spectrum shown in Fig.~\ref{fig:spectrum} indicate the strong rotonization of the collective excitation branch near the phase transition.
By introducing a small fraction of vacancies one can expect the formation of {\it a quadrupolar supersolid} in the strongly interacting regime~\cite{Kurbakov2010}, which is similar to the vacancy-induced Andreev-Lifshitz mechanism~\cite{Andreev1969,Troyer2006,Lutsyshyn2010,Astrakharchik2017,Rota2018}.

\begin{table}[h]\center
\begin{tabular}{l|c|c|c|c}
\hline
pair interaction &$na_s^2$&$\gamma$&$S(k)_{\rm max}$&$n_0/n$\\
\hline\hline
{\bf quadrupoles}&{\bf 1.05}&{\bf 0.269(4)}&{\bf 1.6(1)}&{\bf 0.031(4)}\\
\hline
hard disks~\cite{Xing1990}&0.33&0.279(1)&1.54(2)&---\\
\hline
helium~\cite{prb038002418,prb054006099}&---&0.254(2)&1.7(1)&0.04(1)\\
\hline
dipoles~\cite{Lozovik2007}&2900&0.230(6)&1.7(1)&0.017(6)\\
\hline
Yukawa~\cite{Ceperley1993}&---&0.235(15)&---&---\\
\hline
Coulomb~\cite{Ceperley1994}&---&0.24(1)&---&--- \\
\hline
\end{tabular}
\caption{
Critical values at the gas-solid phase transition in different physical systems:
gas parameter $na_s^2$, 
Lindemann ratio $\gamma$ in crystal phase, 
the height of the first peak in the structure factor $S(k)_{\rm max}$ in the gas phase, 
and the condensate fraction $n_0/n$ in the gas phase.}
\label{table:plug}
\end{table}

It is important to find the properties at the quantum phase transition point. 
In the crystal phase, the value of the Lindemann ratio is found to be equal to $\gamma=0.269(4)$. 
In the gas phase, the height of the first peak in the static structure factor is $S(k)_{\rm max}=1.6(1)$ and the condensate fraction is quite small, $n_0/n=0.031(4)$.
It is instructive to confront the values at the critical point with ones obtained in different 2D bosonic systems. 
Table~\ref{table:plug} summarizes what is known in the literature for other interactions: short-range (hard-disks, helium, Yukawa), extended-range (dipoles) and long-range (Coulomb) ones.
The value of the Lindemann ratio is very similar across all systems, even if the interactions are very different and the order of gas and crystal phases is even inverted in the Coulomb case. 
Also we find that $S(k)_{\rm max}$ and the condensate fraction $n_0/n$ are rather similar in the gas phase at the transition point.
Moreover, our results on the calculation of the condensate fraction of the 2D gas of quadrupoles at $T=0$ are in the quantitative agreement with quantum-field hydrodynamics~\cite{BoronatPRL}.

\begin{figure}
\includegraphics[width=0.75\columnwidth]{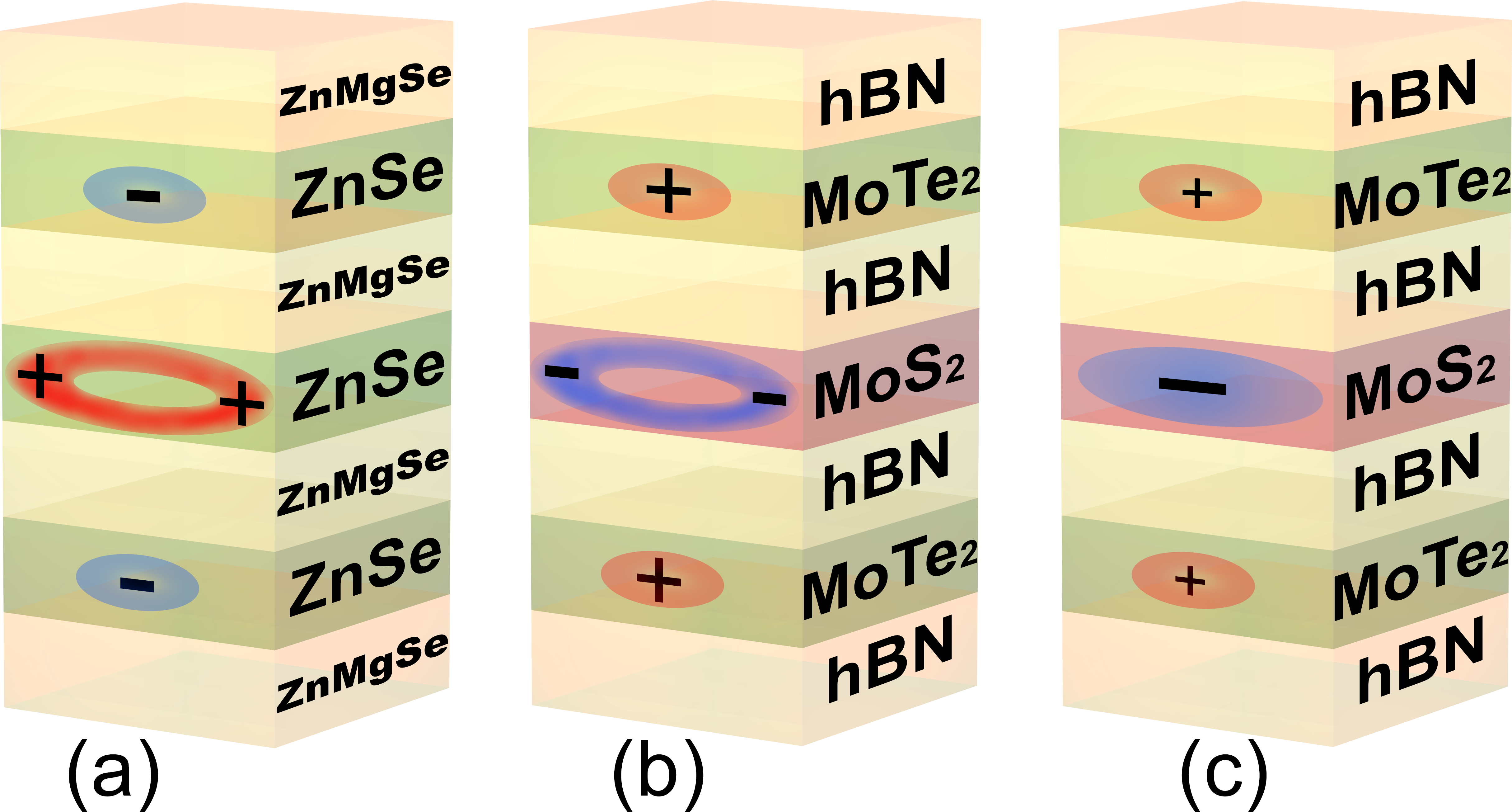}
\vskip -4mm
\caption{Schematic illustration of possible experimental realizations.}
\label{fig:realization}
\end{figure}

As a possible realization of our model, we analyze typical experimental schemes shown in Fig.~5. 
We assume a semiconductor structure consisting of three layers separated by barriers. 
In configuration (a), a quadrupole can be formed out of two holes in the middle layers and two electrons each one in the outer layer. 
Pauli exclusion principle does not allow the holes to be close and their density profile forms a ring.
Assuming a thin ring of radius $R$ and a total charge $-2q$ in the central layer and two point-like $+q$ charges in the lateral layers, the quadrupole moment is equal to
\begin{equation}
Q=3qD^2\sqrt{1+2\alpha(\alpha-1)/3},
\label{Eq:Q}
\end{equation}
where $D$ is the distance between the centers of the central and lateral layers, $\alpha=R^2/D^2$ and the hole charge is $q=e>0$.
Configuration in Fig.~\ref{fig:realization}b is obtained for the specular case with the inverted charges, $q=-e$. 
Configuration in Fig.~\ref{fig:realization}c assumes only a single charge $2q=-e$ in the central layer ($\alpha=0$ in Eq.~(\ref{Eq:Q})) and half-charges $q=+e/2$ in the outer layers.
The physical realizations are based on ZnSe quantum wells\cite{prb057014749} with interlayer separation $D=6$ nm and MoS$_2$/MoTe$_2$ monolayers\cite{prb094155425,prb088085318} with $D=1.667$ nm, according to the schemes shown in Fig.~\ref{fig:realization}. 
The critical densities of the gas-solid phase transition are realistic and correspond to (a) $n_c=1.8\cdot10^{11}$ cm$^{-2}$, (b) $n_c=2.6\cdot10^{12}$ cm$^{-2}$ and (c) $n_c=1.3\cdot10^{13}$ cm$^{-2}$.

\iffalse
\begin{table}[]
\begin{tabular}{|l|l|l|l|}
\hline
Options                    & 1        & 2           & 3           \\ \hline
Central QW                 & WSe$_2$  & MoSe$_2$    & ZnSe        \\ \hline
Side QW                    & MoSe$_2$ & MoSe$_2$    & ZnSe        \\ \hline
Barrier                    & ---      & hBN         & ZnMgSe      \\ \hline
Quaropole                  & $e$$h$   & $e$$e$$h$h$ & $e$$e$$h$h$ \\ \hline
$q/|e|$                    & 0.5      & 1           & -1          \\ \hline
$m/m_0$                    & 1        & 2           & 1.9         \\ \hline
$\epsilon$                 & 10       & 5           & 8.8         \\ \hline
$D$, nm                    & 0.667    & 1.5         & 6           \\ \hline
$R$, nm                    & 0        & 1.5         & 4           \\ \hline
$n_c$, $10^{12}$ cm$^{-2}$ & 250      & 4.2         & 0.22        \\ \hline
$b/D$                      & 1.02     & 3.48        & 3.78        \\ \hline
\end{tabular}
\caption{Three various options for the experimental realization.}
\label{tab:my-table}
\end{table}

\fi

In conclusion, we have obtained the ground-state phase diagram of two-dimensional bosons interacting via quadrupolar potential at zero temperature.
Energetic, structural and coherent properties have been studied in the vicinity of the gas-solid quantum phase transition.
We have demonstrated that the excitation spectrum experiences a strong rotonization in the gas phase close to the critical density. 
We have found an agreement with quantum hydrodynamic calculations for the calculation of the condensate fraction.
Our predictions can be probed in experiments with TMD systems and ultracold gases, where the technique for the observation of roton phenomena recently has been developed. 
Promising candidates for the creation of such phases are quadrupolar excitons in TMD layer structures~\cite{Lukin2020,Rapaport2020},
where the quantum phase transition for the two-component systems has been observed~\cite{Rapaport2020}, and Rydberg atomic ensembles. 

{\it Acknowledgments}. 
G.E.A. acknowledges financial support from the Spanish MINECO (FIS2017-84114-C2-1-P), and from the
Secretaria d'Universitats i Recerca del Departament d'Empresa i Coneixement de la Generalitat de Catalunya  within the ERDF Operational Program of Catalunya (project QuantumCat, Ref.~001-P-001644).
A.K.F. thanks the RSF grant 20-42-05002 for supporting study of the rotonization part. 
I.L.K. and Yu.E.L. are financially supported by the Russian Foundation for Basic Research, grants 19-02-00793 and 20-02-00410.
The authors thankfully acknowledge the computer resources at MareNostrum and the technical support provided by Barcelona Supercomputing Center (RES-FI-2020-3-0011).


\begin{thebibliography}{99}
\bibitem{comment}
The macroscopic value of the Lindemann ratio is obtained in a linear extrapolation in $1/\sqrt N$ according to $\lim_{N\to\infty}\gamma(N)\approx[\sqrt{N_1}\gamma(N_1)-\sqrt{N_2}\gamma(N_2)]/(\sqrt{N_1}-\sqrt{N_2})$ where we consider $N_1 = 34^2=1156$ and $N_2 = 22^2=484$ particles. A similar formula is used for taking the macroscopic limit of the energy per particle $E_0(N)/N$.
\bibitem{Comment1}
We use quantum hydrodynamics for extrapolating the non-condensate fraction $n_0/n$ in the macroscopic limit, see details in Ref.~\cite{BoronatPRL}.
\end{thebibliography}
\end{document}